\documentclass[english,prl,twocolumn,showpacs,amsmath,amssymb,aps,superscriptaddress]{revtex4}
\usepackage[T1]{fontenc}
\usepackage[latin9]{inputenc}
\usepackage{graphicx}
\usepackage{amssymb}

\makeatletter

\makeatletter

\makeatletter

\usepackage{color}

\makeatletter

\usepackage{color}

\makeatletter

\usepackage{bm}

\makeatother

\makeatother

\makeatother

\makeatother

\usepackage{babel}
\makeatother

\begin{document}

\title{Observation of long-lived polariton states in semiconductor microcavities
across the parametric threshold}

\author{D. Ballarini}

\affiliation{Dep. Física de Materiales, Univ. Autonóma de Madrid, 28049 Madrid,
Spain}

\author{D. Sanvitto}

\affiliation{Dep. Física de Materiales, Univ. Autonóma de Madrid, 28049 Madrid,
Spain}

\author{A. Amo}

\affiliation{Dep. Física de Materiales, Univ. Autonóma de Madrid, 28049 Madrid,
Spain}

\altaffiliation{Present address: Laboratoire Kastler Brossel, Université Paris 6, Ecole Normale Supérieure et CNRS, UPMC Case 74, 4 place Jussieu, 75252 Paris Cedex 05, France}

\author{L. Viña}

\affiliation{Dep. Física de Materiales, Univ. Autonóma de Madrid, 28049 Madrid,
Spain}

\author{M. Wouters}

\affiliation{Institute of Theoretical Physics, Ecole Polytechnique Fédérale de
Lausanne EPFL, CH-1015 Lausanne, Switzerland}

\author{I. Carusotto}

\affiliation{BEC-CNR-INFM and Dipartimento di Fisica, Università di Trento, I-38050
Povo, Italy}

\author{A. Lemaitre}

\affiliation{LPN/CNRS, Route de Nozay, 91460, Marcoussis, France}

\author{J. Bloch}

\affiliation{LPN/CNRS, Route de Nozay, 91460, Marcoussis, France}

\begin{abstract}
The excitation spectrum around the pump-only stationary state of a
polariton optical parametric oscillator (OPO) in semiconductor microcavities
is investigated by time-resolved photoluminescence. The response to
a weak pulsed perturbation in the vicinity of the idler mode is directly
related to the lifetime of the elementary excitations. A dramatic
increase of the lifetime is observed for a pump intensity approaching
and exceeding the OPO threshold. The observations can be explained
in terms of a critical slowing down of the dynamics upon approaching
the threshold and the following onset of the soft Goldstone mode. 
\end{abstract}
\maketitle
One of the most striking consequences of the quantum nature of matter
is Bose-Einstein condensation, a phase transition that does not depend
on interactions between the microscopic constituents but is driven
by their indistinguishability. A quite general property of Bose-Einstein
condensed quantum fluids is superfluidity, i.e. the ability to flow
through a container almost without any friction. Since the first discovery
in 4-Helium, this effect was investigated in a variety of quantum
fluids, ranging from 3-Helium to ultracold atomic gases~\citet{BECbooks}.
Inspired by recent developments in the study of phase transitions
in non equilibrium systems~\citet{noneq_phtr}, researchers are presently
working on the extension of the quantum fluid concept to many particle
systems whose state is no longer determined by a thermodynamical equilibrium
condition, but rather by a balance between the external driving and
dissipation.

In this perspective, exciton-polaritons in semiconductor microstructures
appear as very promising objects, as their mixed excitonic and photonic
nature allows for a number of remarkable properties, e.g. a very light
mass, significant interactions, and the possibility of all-optical
manipulation and diagnostics~\citet{review_polar}. On the other
hand, the finite lifetime of polaritons requires some pumping mechanism
to continuously replenish the system, which forbids the establishment
of a true thermal equilibrium in the gas. Several groups have reported
the observation of spontaneous coherence in spatially extended systems
of polaritons far from thermodynamical equilibrium with mechanisms
that can be interpreted as non-equilibrium analogs of Bose-Einstein
condensation (BEC)~\citet{review_coherence}.

In spite of the different pumping schemes (resonant for the OPO case~\citet{OPOs,Huynh2003,OPO_th,Kavokin},
non-resonant for the so-called polariton BEC case~\citet{polar_BEC}),
a $U(1)$ symmetry is spontaneously broken in all cases and coherence
is not simply inherited from the pump beam. Away from the bistability
regime~\citet{noi_param}, the OPO transition is smooth and shows
a critical behavior that closely resembles the one of a second-order
like phase transition~\citet{noi_QMC}. A typical feature of second
order phase transitions is the critical slowing down of the decay
rate of the elementary excitations upon approaching the transition
from below. When a continuous symmetry is spontaneously broken above
the transition point, the lifetime of elementary excitations remains
divergent in the long wavelength limit in agreement with the Goldstone
theorem of statistical mechanics~\citet{huang,pattern}.

In the present Letter, we experimentally investigate and theoretically
model the lifetime of the elementary excitations of a quantum fluid
of polaritons as the pump intensity is spanned across the threshold
in the optical parametric oscillator (OPO) configuration. The steady
state of the system is probed by injecting extra polaritons by means
of a weak pulsed beam, and the decay time of the response is measured
as a function of the pump intensity. A dramatic slowing down of the
dynamics is observed as the threshold is approached from below: close
to the threshold, the decay time can become orders of magnitude longer
than the typical life time of polaritons, and it remains very long
even well above the threshold. Good agreement between the experimental
observations and the theoretical model based on the generalized polariton
Gross-Pitaevskii equation is found, which supports the present theoretical
understanding of the strongly modified dispersion of the elementary
excitations in presence of a strong pump beam. In particular, this
observation suggests the possibility of investigating the polariton
dynamics beyond the limits imposed by the intrinsic polariton life
time.

The microcavity sample used in these experiments is a GaAs/AlAs-based
$\lambda$/2 cavity with a top (bottom) Bragg mirror of 15 (25) Al$_{0.1}$Ga$_{0.9}$As/AlAs
pairs, grown on a GaAs substrate. A 20 nm wide GaAs quantum well is
embedded at the antinode position of the cavity mode. The sample is
kept at a constant temperature of 10 K. The quantum well excitons
are in strong coupling with the cavity mode, with a Rabi splitting
of $2\Omega_{R}=4.4$~meV. All the experiments reported here are
performed in the resonance region, where the $k=0$ cavity-mode has
the same energy as the exciton state.

In the OPO configuration, polaritons are coherently injected into
the microcavity by a pump beam, which resonantly populates a polariton
mode with a defined momentum and energy (pump state). Our pump beam
is a continuous-wave laser source (Ti:Al$_{2}$O$_{3}$), which excites
\textcolor{black}{the sample with an incident angle of $10^{\circ}$
and has a 45}~$\mu$m spot diameter. Its frequency is chosen close
to resonance with the lower polariton branch (LPB) in a way to inject
polaritons with a given wave vector $k_{p}$ and energy $\hbar\omega_{p}$.
Polariton-polariton binary collisions are responsible for the parametric
scattering of pump polaritons into a pair of distinct signal and idler
modes. The efficiency of the parametric processes was optimized by
tuning the pump at a frequency $\hbar\omega_{P}=1.5273$~eV, slightly
above the linear-regime LP dispersion~\citet{whittaker2005}\citet{noi_param},
$\varepsilon_{LPB}(k_{P})$, $\hbar\omega_{p}-\varepsilon_{LPB}(k_{P})\sim\gamma(k_{p})$.
Here $\gamma(k_{p})\sim0.4$~meV is the LPB linewidth at $k_{p}$,
while the laser linewidth is around 0.1~meV. An additional 2 ps-long
{\em probe} pulse, coming from a different laser source with a
repetition rate of 82 MHz, is incident on the sample with a tunable
angle and is focused within the pump spot with a smaller, 25~$\mu$m,
spot diameter. Photoluminescence (PL) is collected and analyzed by
a spectrograph coupled either to a streak- or a conventional CCD-camera.
The emission in the far field is visualized by means of a lens on
the Fourier plane $\left(k_{x};k_{y}\right)$: a direction $k_{y}$
of this plane is selected and energy-resolved by the spectrograph,
which allows the direct observation on the CCD camera of any two-dimensional
section $\left(k_{x};E\right)$ of the polariton dispersion. To obtain
time-resolved PL images a streak camera is used in either a $\left(E;t\right)$
or a $\left(k_{x};t\right)$ configuration, which allows for a complete
study of the dynamics at different energies and momenta.

\begin{figure}
\includegraphics[bb=35bp 145bp 555bp 700bp,clip,width=1\columnwidth]{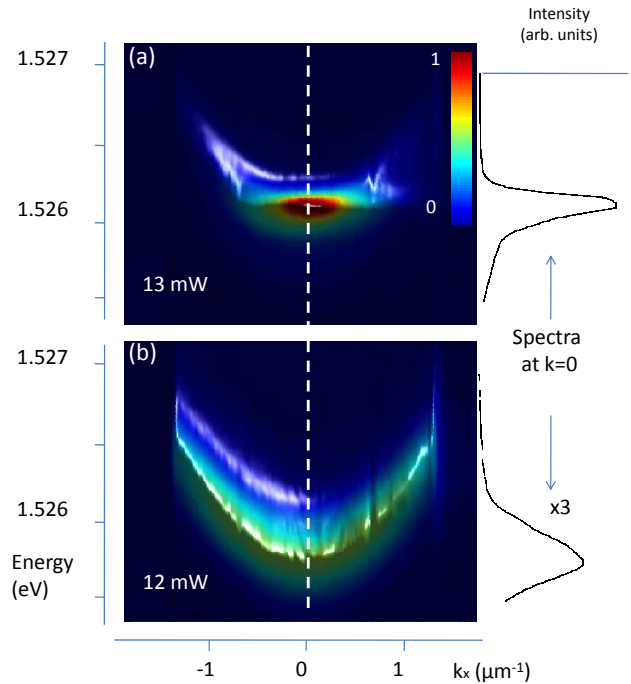}

\caption{Direct experimental observation on the CCD camera of a two-dimensional
section $\left(k_{x};E\right)$, with $k_{y}=0$, of the lower polariton
dispersion for pump powers $I_{p}$ (a) just above (13 mW) and (b)
below (12 mW) the OPO threshold $I_{p}^{th}=12.5$ mW. Energy spectra
at $k_{x}=0$ (dashed line in the figure) are depicted on the right,
with a magnification of a factor 3 for the lower panel of the figure.
The PL emission is normalised to 1 and plotted in a linear color scale.
(color online) \label{fig:distrib} }

\end{figure}

Firstly, we have investigated the stationary-state polariton emission
in the absence of the probe. Typical energy-momentum emission patterns
are shown in Fig.\ref{fig:distrib} for two different values of the
CW pump power $I_{p}$. Although the pump wavevector lies well outside
the $k$-space region imaged in Fig.\ref{fig:distrib}, a small polariton
occupation of the LPB bottom still appears as a consequence of incoherent
relaxation processes even at low pump powers (lower panel). As the
polariton density is very low, the photoluminescence spectrum is concentrated
on the linear-regime LPB branch.

At higher pump intensities, polariton-polariton interactions are able
to significantly modify the emission pattern and, in particular, are
responsible for parametric processes, where two pump polaritons at
$k_{p}$ are transformed into a pair of signal and idler polaritons
of wavevectors $k_{s,i}$, respectively. Thanks to the large size
of the pump spot, the wavevector is approximately conserved, $k_{s}\simeq2\, k_{p}-k_{i}$
and the unique dispersion of polaritons allows for this process to
take place in a resonant way, $2\varepsilon_{LPB}(k_{p})\simeq\varepsilon_{LPB}(k_{s})+\varepsilon_{LPB}(k_{i})$.
The onset of parametric oscillation is clearly visible in the emission
pattern for pump intensities above the threshold $I_{p}^{th}=12.5$
mW (upper panel of Fig.\ref{fig:distrib}): the occupation of the
signal at $k_{s}\simeq0$ becomes in this case very large and the
linewidth of the emission in energy is substantially reduced as compared
to the bare LPB linewidth shown in the lower panel. The spectral narrowing
in energy is accompanied by a significant broadening of the $k$-space
emission. The flat shape of the coherent OPO emission in the $\left(k_{x};E\right)$
plane is however more likely to be a consequence of the peculiar shape
of finite-size non-equilibrium condensates discussed in~\citet{noi_spatial}
rather than an evidence of the diffusive nature of the Goldstone mode~\citet{noi_goldstone}.

\begin{figure}
\includegraphics[bb=20bp 245bp 565bp 585bp,clip,width=1\columnwidth,height=5.1cm]{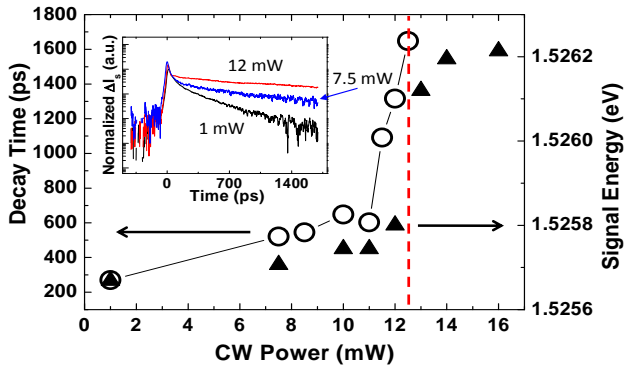}

\caption{Energy (full triangles) and decay time (open circles) of the signal
versus pump power. The red dashed line indicates the pump threshold.
Inset: time evolution of the energy-integrated signal emission $\Delta I_{s}$
for three pump powers (1 mW, 7.5 mW, 12 mW). (color online) \label{fig:I_p}}

\end{figure}

To clearly identify the threshold, we have studied the energy of the
signal emission as a function of pump power (full triangles in Fig.\ref{fig:I_p}).
A smooth and almost linear blue-shift of the signal energy is visible
at low pump powers, while a sudden jump appears for $I_{p}$ just
above 12 mW due to the onset of parametric oscillation. Such a discontinuous
behaviour around the threshold was predicted in Ref~\citet{whittaker2005}\citet{noi_param}:
below the threshold, the energy of the (incoherent) parametric luminescence
is indeed fixed by the slightly blue-shifted LPB dispersion, while
above threshold it is determined by a more complex OPO dynamics that
also involves the idler energy. Far above threshold, the blue-shift
saturates.

The main object of the present study is the response of the system
in its stationary state to an additional weak ($\lesssim0.2\,$mW)
probe pulse that impinges on the sample at a large angle of around
$20^{\circ}$, i.e. in the vicinity of the idler. The evolution of
the system in response to the probe pulse is monitored by investigating
the time- and momentum-resolved signal emission and, in particular,
its decay time. As an example, we have traced in the inset of Fig.\ref{fig:I_p}
the time-evolution of the difference $\Delta I_{s}=I_{s}(pump+probe)-I_{s}(pump)$
between the signal emission intensity in the presence and in the absence
of the probe, respectively, for three different pump intensity values.
To rule out non-linear effects and ensure we are in a linear-response
regime with respect to the probe intensity, we have checked that the
physics of interest is independent of the probe pulse intensity: while
some changes remain visible in the short-time dynamics, the long-time
dynamics simply shows a global rescaling of the observed intensity.

Right after the arrival of the probe pulse, parametric scattering
of pump polaritons into the $k_{s}$ state is stimulated by the small
population of the new idler polaritons injected by the probe: in the
plotted curves, this corresponds to a fast switch-on of the $\Delta I_{s}$
at the probe arrival time. The fast decay on a 30~ps scale is then
followed by a much slower exponential decay on a time scale in the
100 ps range, i.e. orders of magnitude longer than both the empty
cavity decay time (2 ps) and the polariton-polariton scattering time~\citet{Huynh2003},
but still significantly shorter than the probe repetition time of
approximately 12 ns).

As easily seen by comparing the three curves shown in the inset, the
response of the system strongly depends on the intensity $I_{p}$
of the cw pump. While the decay time of the transient decreases for
increasing pump powers and eventually goes below the streak-camera
resolution of $\backsim\,$30~ps, the long decay-time significantly
increases with $I_{p}$. This latter dependence is summarized by the
open circles in the main panel of Fig.\ref{fig:I_p}. The decay time
shows a divergent behavior~\citet{elem_exc_baumberg} for pump powers
approaching $I_{p}=12.5\,\textrm{mW}$; for higher pump powers, it
exceeds the time window of our setup. The parametric nature of the
enhanced life time is confirmed by the coincidence of the divergence
with the signal emission energy jump and the frequency narrowing of
the luminescence (see Fig. \ref{fig:distrib}).

\begin{figure}
\includegraphics[bb=54bp 245bp 515bp 540bp,clip,width=1\columnwidth,height=5.2cm]{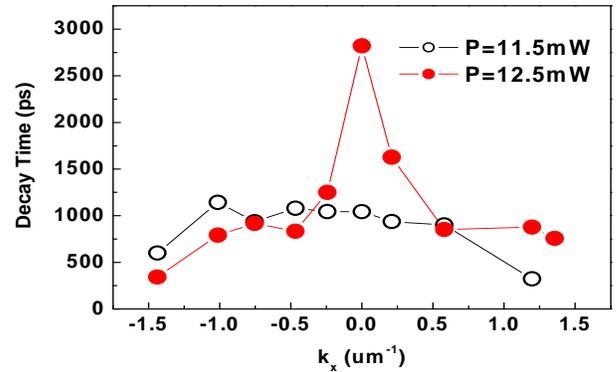}

\caption{$k_{x}$ dependence of the signal decay time for two different pump
powers, 11.5 mW (open circles) and 12.5 mW (full circles). The high
values of the decay time ($>500\,$ps) far from $k_{x}=0$ are due
to the integration over all the spot area and over all the emitted
energies: the intensity profile (not shown) for both 11.5 mW and 12.5
mW is strongly peaked in the parametric scattering region around the
signal state at $k_{x}\simeq0$, while a much weaker and slower incoherent
emission coming from higher energy states (e.g. residual excitons)
and from the border of the spot is responsible for the tails at $\left|k_{x}\right|>0.5\,$$\mu\textrm{m}^{-1}$.
(color online) \label{fig:momentum_resolved} }

\end{figure}

The role of the parametric processes is further evidenced by the momentum-resolved
data shown in Fig.\ref{fig:momentum_resolved}, in which the decay
time of the different $k_{s}$ components of the signal emission is
plotted as a function of the wavevector $k_{x}$. The considered wavevector
range is centered around the value $k_{s}$ where signal emission
would appear if the pump intensity was above threshold. While the
decay time is a smooth function of $k_{x}$ for $I_{p}$ well below
the parametric oscillation threshold (open circles), a marked peak
is apparent in the vicinity of $k_{s}$ ($k_{x}\approx0$) for pump
intensities around and above the threshold (full circles).

A convenient way to interpret the observed slowing-down of the response
to the probe is to use the coherent polariton model based on a pair
of coupled Gross-Pitaevskii-like nonlinear wave equations for respectively
the photon and exciton fields $\psi_{C,X}(\mathbf{r},t)$~\citet{ciuti_review,MI_fest},
\begin{eqnarray}
i\,\frac{\partial\psi_{C}}{\partial t} & = & \left(\omega_{C}(-i\nabla_{\mathbf{r}})-i\frac{\gamma_{C}}{2}\right)\psi_{C}+\Omega_{R}\,\psi_{X}+F(\mathbf{r},t)\\
i\,\frac{\partial\psi_{X}}{\partial t} & = & \left(\omega_{X}\,\psi_{X}-i\frac{\gamma_{X}}{2}\right)\psi_{X}+\Omega_{R}\,\psi_{C}+g\,|\psi_{X}|^{2}\,\psi_{X}.\end{eqnarray}
 We follow the dynamics of the system starting from the $\psi_{X,C}(\mathbf{r},t)=0$
vacuum state. $\omega_{C}(\mathbf{k})$ is the photon dispersion,
while the exciton dispersion is assumed to be flat at $\omega_{X}$.
$\gamma_{C,X}$ are the decay rates of the cavity-photon and the exciton,
respectively. The exciton-exciton interactions are characterized by
the nonlinear coupling coefficient $g$ and $\Omega_{R}$ is the exciton-photon
Rabi coupling. The driving $F(\mathbf{r},t)$ is proportional to the
incident electromagnetic field and has to include both the CW pump
and the pulse probe: once the system has attained its stationary-state
under the CW-pump only, an additional short probe pulse is applied
close to resonance with the idler. The following response of the system
is monitored on the most relevant observables, in particular the polariton
distribution in $k$-space. For the sake of simplicity, we have limited
ourselves to the case of a plane-wave pump with a well-defined wavevector
$k_{p}$ and periodic boundary conditions, while the finite spatial
size of the probe beam is fully taken into account.

\begin{figure}
\includegraphics[width=1\columnwidth]{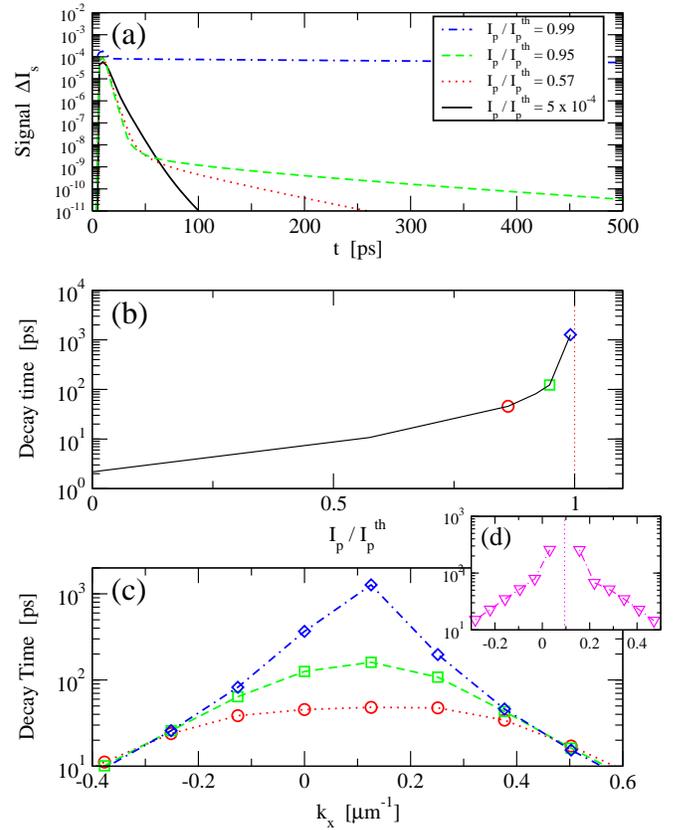}

\caption{Results of numerical calculations. Upper (a) panel: time dependence
of $k$-integrated signal emission $\Delta I_{s}$ for different values
of $I_{p}$; integration is performed over the $k$-space region surrounding
$k_{s}$. Central (b) panel: decay time of $\Delta I_{s}$ as a function
of CW pump intensity. Lower (c) panel: $k$-dependence of decay rate
for different values of $I_{p}/I_{p}^{th}=0.57$, $0.95$, $0.99$
below threshold. Inset (d): $k$-dependence of the decay time for
$I_{p}/I_{p}^{th}=1.15$, above threshold; the vertical dotted line
indicates the coherent signal emission wavevector $k_{s}$, at which
the decay time is infinite by definition of spontaneous symmetry breaking.
(color online) \label{fig:figura_combinata}}

\end{figure}

As discussed in Ref.~\citet{noi_param,MI_fest}, the approach to
the OPO threshold from below is signalled by the decay rate of some
mode tending to zero. As a function of $k$, the decay time results
then strongly peaked around the point where the decay rate is the
smallest: the closer to threshold, the higher the peak value. These
general claims are perfectly visible in the numerical result plotted
in Fig.\ref{fig:figura_combinata}. For pump intensities just below
$I_{p}^{th}$, the evolution of the ntegrated signal emission $\Delta I_{s}(t)$
after the arrival of the probe pulse is characterized by a short transient
followed by a much slower exponential decay, with a time constant
that dramatically increases as the threshold is approached {[}panel
(a)]. By comparing the overall decay time of the integrated $\Delta I_{s}(t)$
{[}panel (b)] with the $k$-dependent decay time {[}panel (c)], it
is immediate to see that the former is determined by the decay time
of the longest-lived mode, a quantity that increases in magnitude
and becomes progressively more peaked as the threshold is approached.

This theoretical result is in good agreement with the experimental
observations for pump intensities in the vicinity of the threshold,
but some specific attention has to be paid at the experimental data
for very low pump power. In this regime, the theoretical calculations
predict that the decay time should go back to the bare polariton lifetime,
while a quite long decay time is observed in the experiment even well
below the oscillation threshold and at wavevectors far from the signal
emission. To explain this behaviour, one can mention the spatial inhomogeneity
of the system that smoothens the $k$-space distribution, as well
as the presence of residual excitons that have accumulated into long-lived
states and that relax down on a long time scale. Clearly, these incoherent
scattering processes are most important for low pump powers, while
coherent parametric processes take it over as the threshold is approached.

As we already mentioned, the decay time above threshold is too long
to be quantitative measured with the present setup. Numerical simulations
do not suffer from such a difficulty, and we summarize here the main
features that one expects for this regime. As a consequence of the
spontaneously broken $U(1)$ symmetry, the spectrum of the elementary
excitations is characterized by the presence of a soft Goldstone mode:
as the wavevector $q=k-k_{s}$ of the excitation tends to zero, both
its frequency and decay rate tend to zero~\citet{noi_goldstone}.
This prediction is confirmed by the numerical results for the $k$-dependent
decay rate that we show in Fig.\ref{fig:figura_combinata}(d): once
again, the peaked structure of the decay rate as a function of $k-k_{s}$
is apparent. It should be mentioned that, even though the long decay
time of the slowest decaying mode eventually appears in the very long
time behaviour of the integrated $\Delta I_{s}(t)$, this feature
is often masked by other, faster decaying branches that contribute
in a much more substantial way to the observed $\Delta I_{s}$ signal.

In conclusion, we have investigated the response of a continuously
pumped optical parametric oscillator to an additional weak pulse in
the vicinity of the idler. The emission from the signal is resolved
in time and its momentum distribution is analyzed as a function of
the pump intensity: when the pump intensity is close or above the
parametric oscillation threshold, a lifetime orders of magnitude longer
than the intrinsic polariton lifetime is observed for polariton modes
in the vicinity of the signal emission. The experimental observations
are explained in terms of a critical slowing down of the elementary
excitation dynamics as the threshold is approached, and then of the
presence of a soft Goldstone mode above the threshold.

This work was partially supported by the Spanish MEC (MAT2005-01388,
NAN2004-09109-C04-04 \& QOIT-CSD2006-00019) and the CAM (S-0505/ESP-0200).
D.B. acknowledges a scholarship of the FPU program of the Spanish
MEC and D.S. thanks the Ramon y Cajal Program. I.C. is indebted to
C. Ciuti for continuous discussions and acknowledges support from
the italian MIUR, the french CNRS and IFRAF and the EuroQUAM-FerMix
program.

\end{document}